\documentclass[12pt,preprint]{aastex}
\usepackage{graphicx,natbib}
\usepackage[percent]{overpic}
\usepackage[usenames, dvipsnames]{color}

\slugcomment{}

\shorttitle{Modeling 13-Dec-2006 CME}
\shortauthors{Fan}

\begin{document}

\title{Modeling the initiation of the 2006 December 13 coronal mass ejection in AR 10930: the structure and dynamics of the erupting flux rope}

\author{Yuhong Fan}
\affil{High Altitude Observatory, National Center for Atmospheric Research, 3080 Center Green Drive, Boulder, CO 80301, USA}

\email{yfan@ucar.edu}

\begin{abstract}
We carry out a three-dimensional magneto-hydrodynamic (MHD)
simulation to model the initiation of the
coronal mass ejection (CME) on 13 December 2006 in the emerging $\delta$-sunspot
active region NOAA 10930.  The setup of the simulation is similar to a
previous simulation by Fan (2011), but with a significantly widened simulation
domain to accommodate the wide CME. The simulation shows that the CME can result
from the emergence of a east-west oriented twisted flux rope whose positive,
following emerging pole corresponds to the observed positive rotating
sunspot emerging against the southern edge of the dominant pre-existing
negative sunspot. The erupting flux rope in the simulation accelerates
to a terminal speed that exceeds 1500 km/s and undergoes a counter-clockwise
rotation of nearly $180^{\circ}$ such that its front and flanks all exhibit
southward directed magnetic fields, explaining the observed southward magnetic
field in the magnetic cloud impacting the Earth. With continued driving of
flux emergence, the source region coronal magnetic field also shows the
reformation of a coronal flux rope underlying the flare current sheet of the
erupting flux rope, ready for a second eruption. This may explain the build
up for another X-class eruptive flare that occurred the following day from
the same region.

\end{abstract}


\section{Introduction \label{sec:intro}}
MHD modeling of the realistic coronal magnetic field evolution of CME events
is critically important for understanding the connection between CMEs and
interplanetary magnetic clouds, and determining/predicting their
geo-effectiveness in the resulting space weather events
\citep[e.g][]{Mikic:etal:2008,Titov:etal:2008,Kataoka:etal:2009,
Downs:etal:2015}.
The 13 December 2006 eruptive event from active region (AR) 10930
produced an X3.4 class flare and a fast, halo CME with an estimated speed
of at least $1780$ km/s \cite[e.g.][]{Liu:etal:2008,Ravindra:etal:2010}.
The CME evolved into an interplanetary magnetic cloud and reached the
Earth on 2006 December 14-15, with a strong prolonged southward
directed magnetic field in the magnetic cloud, causing a major
geomagnetic storm \citep[e.g.][]{Liu:etal:2008,Kataoka:etal:2009}.
The photospheric magnetic field evolution of AR 10930 was well
observed by the Solar Optical Telescope (SOT) on board the Hinode
satellite over a period of several days before, during, and
after the eruption.
Many analyses of the observed magnetic flux emergence, buildup
of current and free magnetic energy, and changes of the photospheric fields
associated with the X-class flare have been carried out
\citep[e.g.][]{Kubo:etal:2007,Zhang:etal:2007,Schrijver:etal:2008,Min:Chae:2009,
Gosain:etal:2009,Ravindra:etal:2010,Ravindra:etal:2011b}.
The evolution of AR 10930 was characterized
by an emerging $\delta$-sunspot with a growing positive polarity spot
against the south edge of a dominant pre-existing negative spot,
displaying substantial counter-clockwise rotation and eastward motion
as it grew \citep[e.g.][]{Kubo:etal:2007, Min:Chae:2009}.
The growth of the positive rotating sunspot is accompanied by
the emergence of fragmented negative polarity pores to the west
of the emerging positive spot, suggesting that they are the
counterparts of an east-west oriented emerging bipolar pair
\citep{Min:Chae:2009}.

\citet[][hereafter F11]{Fan:2011} has carried out an MHD simulation
of the magnetic field evolution associated with the onset of the
eruptive flare and CME in AR 10930 on 2006 December 13.
Motivated by the observed photospheric magnetic flux emergence
pattern, the simulation assumes the emergence of an
east-west oriented magnetic flux rope into a pre-existing coronal
magnetic field constructed based on the Solar and Heliospheric
Observatory (SOHO)/Michelson Doppler Imager (MDI) full-disk
magnetogram of the photospheric magnetic field at 20:51:01 UT on
December 12.
It is found that several observed features of the CME source
region and the eruptive flare, such as the pre-eruption X-ray
sigmoid, the evolution of the flare ribbons, and the morphology
of the X-ray post-flare loop brightening, can be qualitatively
explained by the modeled coronal magnetic field evolution.
The simulation results in the eruption of a coronal flux rope
that reaches a terminal speed of about 830 km/s and shows
significant writhing or rotation as it erupts.  However,
because of the narrow simulation domain used (about
$30^{\circ}$ in latitudinal width), the erupting flux rope
that expands rapidly
becomes severely constrained by the side boundary walls
almost immediately
after the onset of the eruption and its acceleration and
rotation are significantly impacted.
As can be seen in Figure 7 of F11, the erupting flux rope begins
to hit the south boundary wall soon after the onset of eruption
when the front of the cavity has only just reached about 1.3 solar
radii. The subsequent acceleration and evolution (including the expansion
and rotation) of the flux rope are then severely impeded and constrained
by the side walls and cannot be properly followed.
In this work, we improve upon the simulation of F11 by
significantly widening the simulation domain (to about
$117^{\circ}$ in latitudinal width and about $98^{\circ}$ in
longitudinal width), and inclusion of a much more extended
region of the observed photospheric normal flux distribution
in the construction of the pre-existing coronal potential
field.
The latter also provides a more accurate description of the
decline profile with height of the confining potential magnetic field,
which would affect the development of the torus instability and the
acceleration of the flux rope in the lower corona
\citep[e.g.][]{Toeroek:Kliem:2007}.
We found that the resulting erupting flux rope accelerates
to a higher terminal speed of about 1500 km/s.
It undergoes a counter-clockwise rotation (as viewed from
above) of about $180^{\circ}$, such that both the front
and the flanks of the final expanding flux rope  
are showing southward directed magnetic field, opposite to
the field direction for the top of the pre-eruption flux rope
and the its overlying potential field.

\section{The Numerical Model \label{sec:model}}
For the simulation carried out in this study, we solve the MHD equations
in spherical geometry as given in \citet[][hereafter F12]{Fan:2012}.
We assume an ideal gas with a low adiabatic index
$\gamma = 1.1$ for the coronal plasma, which allows it to maintain its high
temperature without an explicit coronal heating. The MHD equations are solved
numerically with the MFE code described in F12.  No explicit viscosity and
magnetic diffusion are included in the momentum and the induction equations.
However numerical dissipations are present at regions of sharp gradient and
the non-adiabatic heating due to numerical dissipation is implicitly
converted into the internal energy by solving the total energy equation
in conservative form. Compared to the simulation of F11, the difference in
the formulation is that here we also include the field-aligned thermal
conduction in the energy equation as given in F12, which was not included in
the previous simulation of F11.

\begin{figure}
\centering
\includegraphics[width=0.75\textwidth]{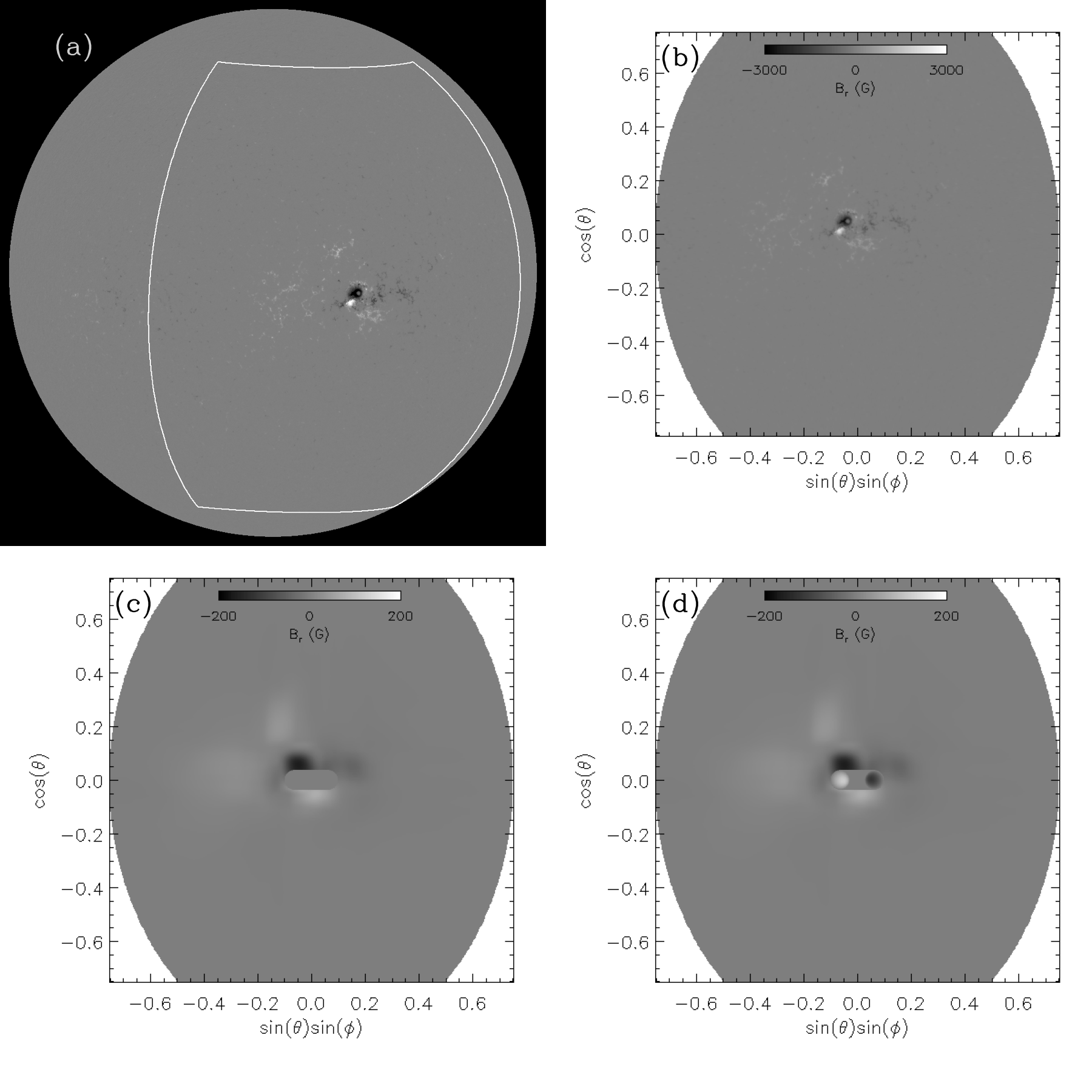}
\caption{(a) SOHO MDI full-disk magnetogram at 20:51:01 UT on December 12, 2006.
The white box encloses the area to be used as the lower boundary
of the simulation domain. (b) Radial magnetic field $B_r$
in the region enclosed by the white box in (a) as viewed
straight-on at the center of the region. (c) $B_r$ on the lower boundary of the
simulation domain after a Gaussian smoothing has been applied and with the
field in a central region zeroed out, where the emergence of a twisted magnetic
torus is to be imposed.
(d) $B_r$ on the lower boundary at the end of the simulation.
A movie of the $B_r$ evolution on the lower boundary during the whole course
of the simulation is given in the online version of the article}
\label{fig:initdomain}
\end{figure}

Similar to the simulation setup of F11, we
impose at the lower boundary the emergence of the upper half of a twisted
magnetic torus into a pre-existing coronal potential magnetic field. The
potential field is constructed using the MDI full-disk magnetogram taken at
20:51:01 UT, December 12, 2006. However
here we incorporate a much wider simulation domain compared to F11.  As
shown in Figure \ref{fig:initdomain}(a), from
the MDI full-disk magnetogram, a much wider region centered on the emerging
$\delta$-sunspot, corresponding to the region enclosed by the white box
with a latitudinal width of $117^{\circ}$ and a longitudinal width of $98^{\circ}$, is
extracted to be the lower boundary of the simulation domain.
The simulation domain as described in the simulation spherical coordinate
system is given by $r \in [R_{\odot}, 6.25 R_{\odot}]$,
$\theta \in [31.4^{\circ}, 148.6^{\circ}]$,
and $\phi \in [-49.2^{\circ},49.2^{\circ}]$.
The center of the white-boxed region shown in Figure \ref{fig:initdomain}(a)
is the center of the simulation lower boundary at
$\theta= 90^{\circ}$ and $\phi=0^{\circ}$.
We use a non-uniform grid of $576(r) \times 480 (\theta)
\times 504 (\phi)$ that is stretched in all three dimensions to resolve
the simulation domain.
In $r$, the grid size is 1 Mm for $r < 1.65 R_{\odot}$
and then it increases geometrically, reaching about 92 Mm at the outer
boundary of $6.25 R_{\odot}$. Horizontally in $\theta$ and $\phi$, the
grid size is about $0.00148 \; rad$ (or $0.0848^{\circ}$) in the central
area within $\sim 10^{\circ}$ heliographic distance from the center, and then
increases geometrically to about $0.0162 \; rad$ (or $0.929^{\circ}$)
towards the $\theta$ boundaries, and to about $0.0131 \; rad$
(or $0.748^{\circ}$) towards the $\phi$ boundaries.

For the $\theta$ and $\phi$ boundaries,
we assume non-penetrating, stress-free, and electrically perfect conducting walls.
For the outer $r$ boundary, we use simple outward extrapolations to
allow plasma and magnetic flux to flow through.
At the lower boundary, the normal magnetic field $B_r$ extracted from
the MDI full disk magnetogram as viewed straight on the center of the
region is shown in Figure \ref{fig:initdomain}(b).
As is in F11, we apply a smoothing of $B_r$ using a Gaussian filter
that reduces the peak field strength from about $3000$ G to $164$ G. This smoothing is done
since the lower boundary density is set to be that of the base of the corona
(see below), and therefore a drastic reduction of the field strength from
that measured on the photosphere is made to avoid an extremely high peak
Alfv{\'e}n speed that would severely limit the time step of numerical integration.
With the smoothed $B_r$, we then zero out the magnetic flux in a central area that
roughly encloses the region of the observed flux emergence, including the rotating
positive sunspot and the negative emerging pores to the west of it, as
shown in Figure \ref{fig:initdomain}(c).
The zeroed out area is the region on the lower boundary where we impose
the emergence of a twisted magnetic torus as described below.
The potential field extrapolated from the normal
magnetic field shown in Figure \ref{fig:initdomain}(c) is used as the
initial coronal magnetic field for the simulation.
For the initial atmosphere we assume a static polytropic atmosphere:
\begin{equation}
\rho = \rho_{R_{\odot}} \left [ 1 - \left ( 1- \frac{1}{\gamma} \right )
\frac{GM_{\odot}}{R_{\odot}} \frac{\rho_{R_{\odot}}}{p_{R_{\odot}}} \left ( 1
- \frac{R_{\odot}}{r} \right ) \right ]^{\frac{1}{\gamma - 1}}
\end{equation}
\begin{equation}
p = p_{R_{\odot}} \left [ 1 - \left ( 1- \frac{1}{\gamma} \right )
\frac{GM_{\odot}}{R_{\odot}} \frac{\rho_{R_{\odot}}}{p_{R_{\odot}}} \left ( 1
- \frac{R_{\odot}}{r} \right ) \right ]^{\frac{\gamma}{\gamma - 1}}
\end{equation}
where $\rho_{R_{\odot}} = 8.365 \times 10^{-16} $ g ${\rm cm}^{-3}$
and $p_{R_{\odot}} = 0.152$ dyne ${\rm cm}^{-2}$ are the
initial density and pressure at the lower boundary, with the
temperature at the lower boundary being 1.1 MK.
Thus for the initial static equilibrium, the peak Alfv{\'e}n speed is
$21,800$ km/s in the main negative sunspot at the lower boundary, and the
sound speed is $141$ km/s at the lower boundary.
The Alfv{\'e}n speed is greater than the sound speed
in most of the simulation domain.

In the area where the flux is zeroed out on the lower boundary surface
($r=R_{\odot}$), we specify the following
time dependent transverse electric field ${\bf E}_{\perp}|_{r=R_{\odot}}$
to drive the kinematic emergence of a twisted magnetic
torus ${\bf B}_{\rm tube}$ at a velocity ${\bf v}_{\rm rise}$:
\begin{equation}
{\bf E}_{\perp}|_{r=R_{\odot}} = {\hat{\bf r}} \times \left [ \left (
- \frac{1}{c} \, {\bf v}_{\rm rise} \times
{\bf B}_{\rm tube} (R_{\odot}, \theta, \phi, t) \right )
\times {\hat{\bf r}} \right ].
\label{eq:emf}
\end{equation}
${\bf B}_{\rm tube}$ is an axisymmetric torus
specified below in its own local spherical polar coordinate system
($r'$, $\theta'$, $\phi'$).
The origin of the ($r'$, $\theta'$, $\phi'$)
coordinate system is the center of the torus, and is
located at ${\bf r} = {\bf r}_c = (r_c, \theta_c, \phi_c)$
in the Sun-centered simulation spherical coordinate system $(r, \theta, \phi)$.
The polar axis of the ($r'$, $\theta'$, $\phi'$)
coordinate system is the symmetric axis of the torus,
and is parallel to the $- {\hat {\bf \theta}}$ direction
at the position ${\bf r}_c$.
In the ($r'$, $\theta'$, $\phi'$) coordinate system, ${\bf B}_{\rm tube}$ is
given by:
\begin{equation}
{\bf B}_{\rm tube} = \nabla \times \left (
\frac{A(r',\theta')}{r' \sin \theta' } \hat{\bf \phi'} \right )
+ B_{\phi'} (r', \theta') \hat{\bf \phi'},
\label{eq:btube}
\end{equation}
\begin{equation}
A(r',\theta') = \frac{1}{4} q a^2 B_t
\left( 1 - \frac{\varpi^2(r',\theta')}{a^2} \right)^2 ,
\label{eq:afunc}
\end{equation}
\begin{equation}
B_{\phi'} (r', \theta') = \frac{a B_t}{r' \sin \theta'}
\left( 1 - \frac{\varpi^2(r',\theta')}{a^2} \right),
\label{eq:bph}
\end{equation}
\begin{equation}
\varpi (r',\theta') = (r'^2 + R'^2 -2r'R' \sin \theta')^{1/2},
\label{eq:varpi}
\end{equation}
where $a=0.035 R_{\odot}$ is the minor radius of the torus,
$R'=0.063 R_{\odot}$ is the major radius of the torus,
$\varpi$ denotes the distance to the curved axis
of the torus,
$q/a=0.082 \; {\rm rad/Mm}$ corresponds to the rate of field
line twist (rad per unit length) about the curved axis of the torus,
and $B_t a/R' = 111$ G is the field strength at the curved axis of
the torus. ${\bf B}_{\rm tube}$ is truncated to zero for $\varpi > a$.
The field line twist rate of the emerging torus used here is about
1.33 times that used in F11.
Initially the center of the ($r'$, $\theta'$, $\phi'$)
coordinate system (i.e. the center of the torus) is located at
${\bf r}_c = (r_c = 0.902 R_{\odot}, \; \theta_c = 90^{\circ}, \; \phi_c = 0^{\circ})$.
Thus the torus
initially lies in the equatorial plane in the simulation coordinate system,
below the lower boundary with its outer edge just touching the lower boundary.
For specifying the time dependent ${\bf E}_{\perp}|_{r=R_{\odot}}$,
we assume that the center of the torus
${\bf r}_c$ is rising at a constant velocity ${\bf v}_{\rm rise} = v_{\rm rise}
{\hat {\bf r}_c}$ with $v_{\rm rise} = 9.75$ km/s
(much smaller than the Alfv{\'e}n speed and the sound speed in the coronal
domain). The velocity field on the lower boundary is
uniformly ${\bf v}_{\rm rise}$ in the area where the emerging torus intersects
the lower boundary and zero outside.
Compared to F11, which used a much larger $v_{\rm rise} = 98$ km/s
in the early phase of emergence and then slowed down to
$v_{\rm rise} =19.5$ km/s when getting closer to the onset of eruption, 
here we use a much slower but constant $v_{\rm rise} = 9.75$ km/s to drive
the emergence continuously.

The resulting ${\bf E}_{\perp}|_{r=R_{\odot}}$ (eq. \ref{eq:emf})
produces the emergence of an east-west oriented bipolar region as shown
in a movie of $B_r$ evolution on the lower boundary given in the online
version of Figure \ref{fig:initdomain}, and Figure \ref{fig:initdomain}(d)
shows a snapshot of $B_r$ on the lower boundary at the end of the simulation.
The imposed flux emergence pattern is qualitatively representative of
the observed flux emergence pattern of the region \citep{Kubo:etal:2007,
Min:Chae:2009}. It is found that the emergence of the positive rotating
sunspot moving eastward is accompanied by scattered pores of negative
polarity emerging and moving westward, suggesting that they are the
counterparts of an east-west oriented emerging bipolar pair
\citep[see Figure 2 in][]{Min:Chae:2009}.
For our imposed flux emergence pattern (Figure \ref{fig:initdomain}(d) and
the associated online movie)
produced by the emergence of a twisted
magnetic torus, the positive spot of the emerging bipolar pair corresponds
to the positive rotating sunspot moving eastward along the southern edge
of the large pre-existing sunspot, and the negative emerging spot represents
the collection of the observed scattered negative emerging pores.

\section{Results \label{sec:results}}
\subsection{Overview of evolution}
\begin{figure}
\centering
\includegraphics[width=0.75\textwidth]{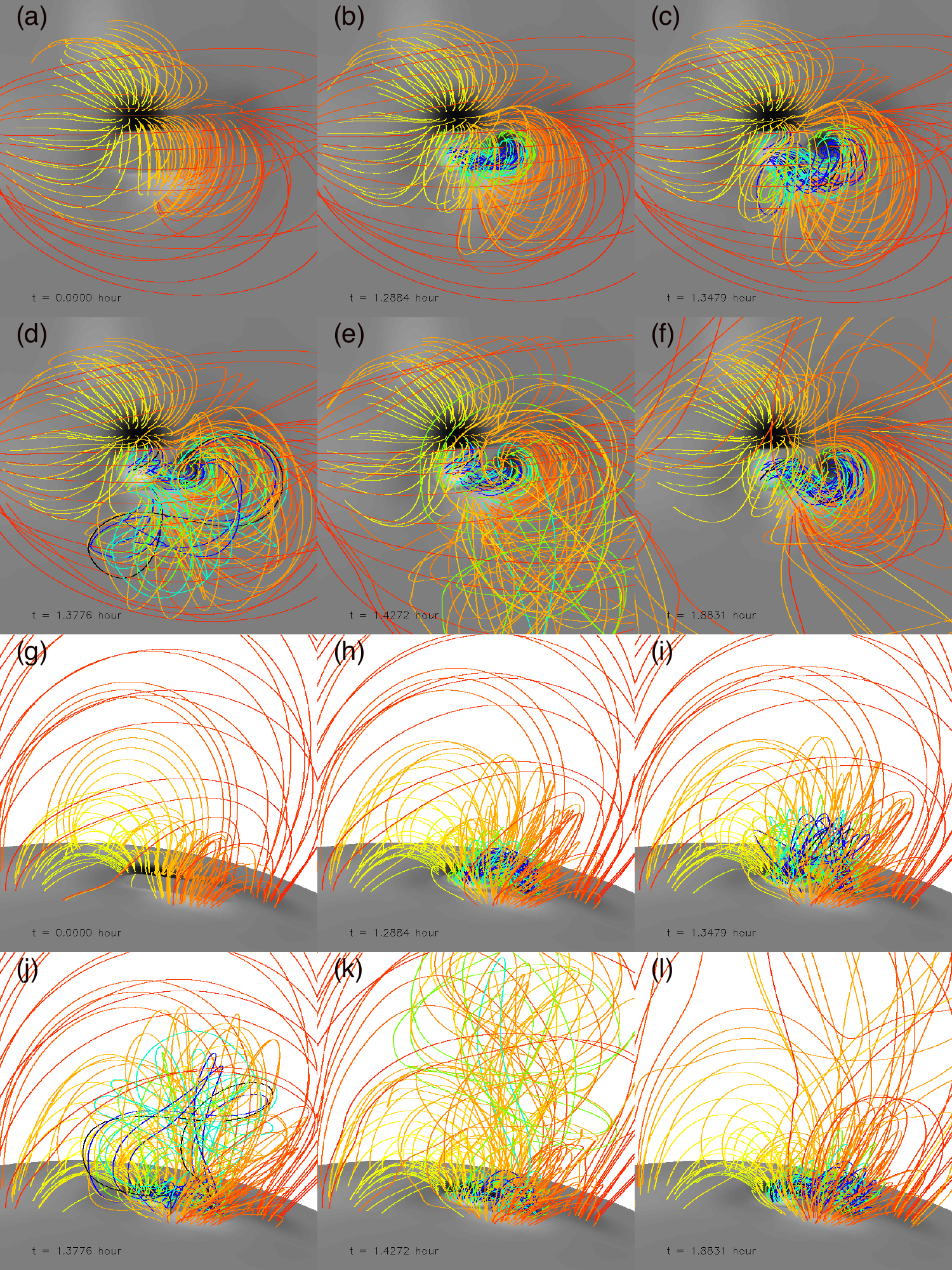}
\caption{Snapshots of the 3-dimensional magnetic field evolution over
the whole course of the simulation, with the top 2 rows showing a
perspective view from the Earth's line-of-sight, and the bottom 2 rows
showing a side view. A movie of the 3D field evolution
is available in the online version.}
\label{fig:fdl3d_ev}
\end{figure}

\begin{figure}
\centering
\includegraphics[width=0.75\textwidth]{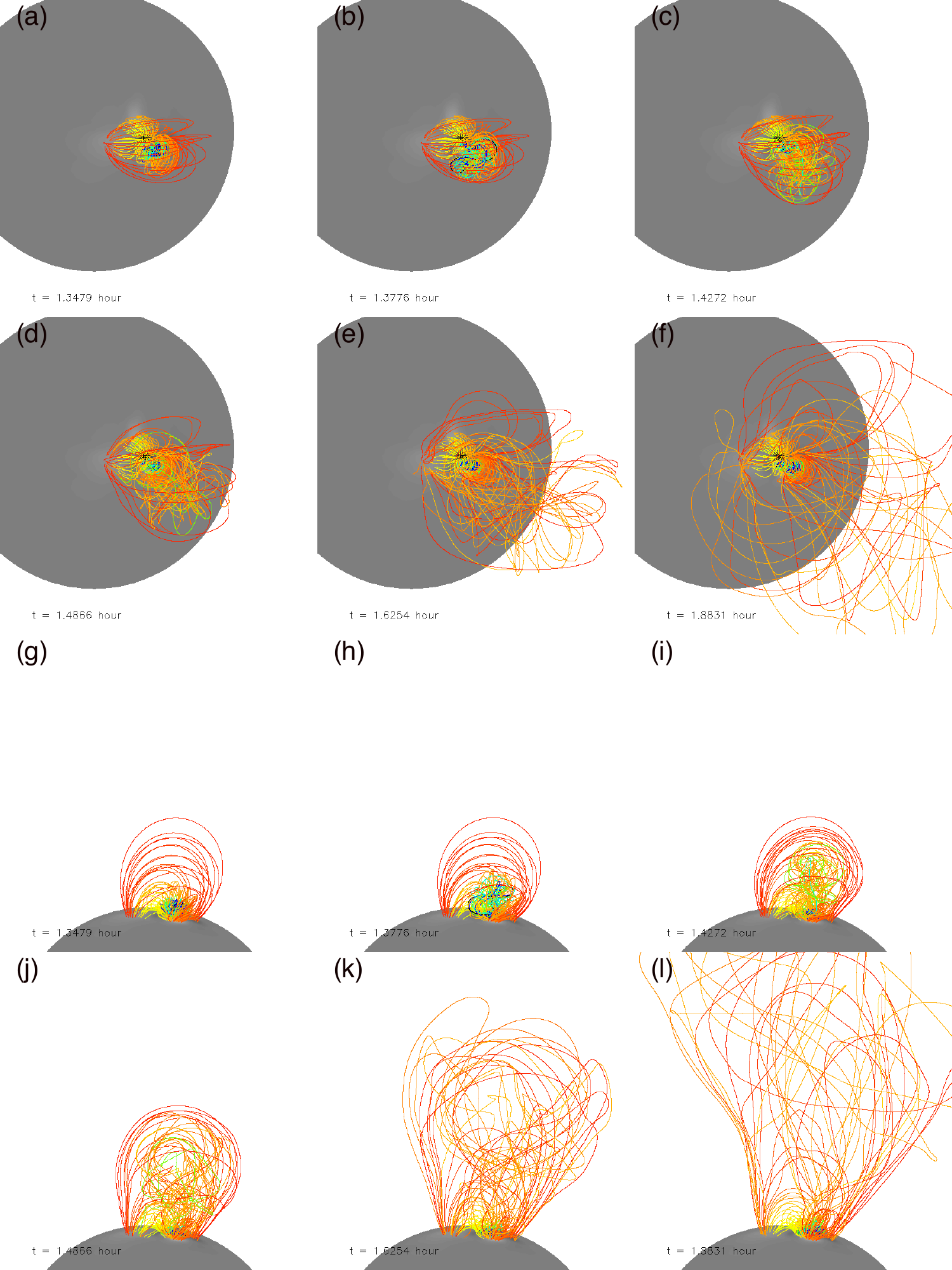}
\caption{Snapshots showing the zoomed out views of the 3D coronal magnetic
field after the onset of the eruption. Top two rows show the perspective
view from the Earth's line-of-sight, and the bottom rows show a zoomed
out side view. A movie of the 3D field evolution 
is available in the online version.}
\label{fig:fdl3dfar_ev}
\end{figure}

Figure \ref{fig:fdl3d_ev} shows snapshots of the evolution of the
3D coronal magnetic field over the whole course of the simulation,
with the top 2 rows of images showing a perspective view from the Earth's
line-of-sight and the bottom 2 rows showing a side view.
Figure \ref{fig:fdl3dfar_ev} shows snapshots of two zoomed-out views
of the 3D field evolution during the later stage of the simulation.
A movie of the 3D field evolution combining the 4 views shown in the
above two figures is available in the online version.
We trace the field lines shown in Figures \ref{fig:fdl3d_ev} and
\ref{fig:fdl3dfar_ev} in the following way. We use a set of fixed
foot points in the ambient field region outside of the emerging flux
region to trace the red, orange, and yellow field lines. For tracing the field
lines (green, blue, and black field lines) from the emerging flux region
on the lower boundary, we track a set of foot points that connect to the
same field lines of the subsurface emerging torus and the field lines are
colored based on the flux surfaces of the subsurface torus.
Figures \ref{fig:fdl3d_ev}(a)(g) show the initial potential magnetic
field, which is constructed using the lower boundary $B_r$ shown in Figure
\ref{fig:initdomain}(c).
With time a twisted coronal flux rope (as represented by the green, blue 
and black field lines in Figures \ref{fig:fdl3d_ev}(b)(h)) is built up
quasi-statically, as a result of the imposed flux emergence at the lower
boundary described in Section \ref{sec:model}.

Figure \ref{fig:emfree_ek_evol} shows the evolution of the free magnetic
energy (top panel), which is the total magnetic energy $E_m$ minus the energy
$E_p$ of the corresponding potential magnetic field extrapolated using the
current normal magnetic field distribution at the lower boundary, and
the evolution of the kinetic energy $E_k$ (bottom panel).
Figure \ref{fig:vr_axis_sheath} shows the rise velocity at the apex of the
axial field line of the coronal flux rope (diamond points),
which connects to the axial
field line of the subsurface emering torus, and the rise velocity at the
front of the erupting coronal flux rope cavity (cross points) in the later
time period.
From $t=0$ to roughly $t=1.3$ hour, the coronal magnetic field evolves
quasi-statically, with the free magnetic energy being built up (top panel
of Figure \ref{fig:emfree_ek_evol}), and with the coronal flux rope emerges
and rises quasi-statically with very small rise velocity
(Figure \ref{fig:vr_axis_sheath}).
The coronal flux rope begins to erupt at about $t=1.3$ hour, where
the free magnetic energy undergoes a significant decrease, the kinetic energy
undergoes a rapid increase (see Figure \ref{fig:emfree_ek_evol}), and the
flux rope axis shows a rapid acceleration (Figure \ref{fig:vr_axis_sheath}).
Figures \ref{fig:fdl3d_ev}(c)(d)(e)(i)(j)(k) (also the associated animation
in the online version of the paper) show that the flux rope field
rooted in the emerging region (green, blue, and black field lines) erupts upward,
pushing the ambient field (red, orange, yellow field lines) outward,
and shows a counter-clockwise rotation as it erupts.
The apex of the axial field line accelerate to about $1300$ km/s (diamond
points in Figure \ref{fig:vr_axis_sheath}) before it reconnects.
Subsequently we track the rise velocity at the front of the erupting
flux rope cavity (cross points in Figure
\ref{fig:vr_axis_sheath}) and find that its rise speed increases to over
$1500$ km/s at the end of the simulation when it reaches about 4.5 $R_{\odot}$.
Figure \ref{fig:vrmeri_dmeri} shows snapshots of the evolution of the radial
velocity $V_r$ in the central meridional plane across the erupting flux
rope (top panels), and the corresponding evolution of the
density in the same meridional plane (bottom panels).
A movie showing the evolution of $V_r$
and density in the meridional plane is also available with the online
version of the figure.
We see the development of a wide ejecta, which consists of a central low
density cavity region (containing the twisted erupting flux rope)
surrounded by a denser sheath (Figures \ref{fig:vrmeri_dmeri}(d)(e)(f)).
Compare Figure \ref{fig:vrmeri_dmeri}(d) here with Figure 7 in F11.
They are both showing
roughly the same stage of the eruption when the front of the flux rope
cavity reaches about 1.3 $R_{\odot}$.  In F11, the erupting flux rope is already
hitting the south boundary wall because of the narrow simulation domain used
there. As a result the subsequent acceleration of the flux rope becomes
severely impeded and the speed measured at the front of the flux rope cavity
becomes saturated at about 830 km/s as shown in Figure 6 of F11.  Here with
the significantly widened simulation domain (see Figures
\ref{fig:vrmeri_dmeri} (d)(e)(f)), we are able to follow the subsequent
acceleration and evolution of the flux rope.
The front of the sheath and the front of the
cavity are found to attain a speed of about $1500$ km/s at the end of the
simulation (see Figures \ref{fig:vrmeri_dmeri}(c)(f)).
The velocity shown by the cross points in Figure \ref{fig:vr_axis_sheath}
is measured at the front edge of the cavity on the $z=0$ line in the
meridional plane shown in Figure \ref{fig:vrmeri_dmeri}.

As the twisted flux rope erupts, it continually reconnects with the ambient
coronal magnetic field, and towards the end of the simulation, the outward
erupting fields have become entirely rooted in the ambient field region
outside of the emerging flux region on the lower boundary. This
can be seen in Figures \ref{fig:fdl3dfar_ev}(e)(f)(k)(l), where the red,
orange, and yellow colors of the erupting field lines indicate that they are rooted
in the ambient field region (see more description later in
Section \ref{sec:erupting_rope}). Furthermore, due to
the continued flux emergence, the free magnetic energy is built up again
after the rapid release of the first eruption
(see top panel of Figure \ref{fig:emfree_ek_evol}) and a new closed
twisted flux rope has reformed connecting the emerging bipolar region,
poised for a second eruption (see the blue and green field lines in Figures
\ref{fig:fdl3d_ev}(f)(l)). From the movie associated with
the online version of Figure \ref{fig:fdl3d_ev}, it can be seen that the
reformed flux rope has begun to accelerate for the second
eruption at the end of the simulation at about $t=1.88$ hour.

\subsection{Onset of the eruption}
At the onset of the eruption at about $t=1.3$ hour, the twist of the field
lines about the axis of the emerged flux rope has reached about $1.2$
winds between their anchored ends.  Therefore the flux rope is close to
the threshold of twist ($2.49 \pi$ or about $1.25$ winds) for the onset of
the helical kink instability \citep[e.g.][]{Hood:Priest:1981}.
The $1.2$ winds (or $432^{\circ}$ rotation) of field line twist in the emerged
flux rope is within the observed range of the total twist transported into the
corona as measured by the rotation of the emerging positive sunspot:
at least $240^{\circ}$ obtained by \citet{Zhang:etal:2007}, and $540^{\circ}$
obtained by \citet{Min:Chae:2009}.
We also show in the bottom panel of Figure \ref{fig:bpdecay_alpha},
$d \ln (B_p) / d \ln (H)$, the decay rate with
height $H$ of the corresponding potential magnetic field $B_p$ along the
central vertical slice in the central meridional plane across the flux rope,
and in the top panel of Figure \ref{fig:bpdecay_alpha}, the profile of
$\alpha \equiv (\nabla \times {\bf B}) \cdot {\bf B} / B^2 $ along the same
vertical slice.  The value of $\alpha$ is a measure of the twist rate of
the magnetic field, and the height range where $\alpha$ is significantly
negative indicates the height range of the flux rope cross-section.
We see from Figure \ref{fig:bpdecay_alpha} that the top of the flux rope
cross-section has reached about $H=0.15 R_{\odot}$ as marked by the vertical
dotted lines and that the top portion of the flux rope
has reached the region where the decay rate of the potential magnetic
field $-d \ln (B_p) / d \ln (H)$ has exceeded the magnitude of 1.5, which is
the threshold for the flux rope to develop the torus instability
\citep[e.g][]{Kliem:Toeroek:2006,Isenberg:Forbes:2007}.
Thus both the helical kink instability and the torus instability are likely
playing a role in triggering the eruption of the coronal flux rope.
We found that with a wider simulation domain and the inclusion of a wider
range of the observed ambient normal flux distribution on the lower boundary,
the magnitude of the decay rate of the potential field is greater compared to that
obtained for F11, in the lower height range from 0 to $0.6 R_{\odot}$.  This
might have contributed to a higher terminal speed
of the erupting flux rope in the present case
\citep[e.g.][]{Toeroek:Kliem:2007}. However, the dominant reason
for the lower saturation speed of the front of the flux rope 
found in F11 is due to the side wall boundary which begins to impede the
acceleration of the flux rope soon after the onset of the eruption (see Figure 7
of F11).

Figure \ref{fig:cfdl_sigmoidsh} shows 3D field lines (colored based on the
temperature) of the coronal magnetic field at a time ($t=1.26$ hour) just
before the onset
of the eruption (panel (a)), compared with the Hinode X-Ray Telescope (XRT)
image of the region just prior to the onset of the flare (panel (b)).
Similar to F11, we find that the central core field of the flux rope is
strongly heated because of the formation of a sigmoid shaped current layer.
In Figure \ref{fig:cfdl_sigmoidsh}(a) we have densely traced these highly
heated field lines, whose temperature has reached about 10 MK.
The heated sigmoid shaped core field may give rise to the
bright sigmoid loops in the pre-eruption region observed in Hinode XRT
images \citep[Figure \ref{fig:cfdl_sigmoidsh}(b), also][see
their Figure 1]{Su:etal:2007}.

After the flare onset, the soft X-ray emission becomes
completely dominated by the brightening of a sigmoid-shaped row of
post-flare loops as shown in a Hinode XRT image in
Figure \ref{fig:hinode_ribbon_xray}(b), and the corresponding
Hinode Solar Optical Telescope (SOT) observation
(Figure \ref{fig:hinode_ribbon_xray}(a)) shows the two-ribbon
brightening in the lower atmosphere.
The two bright ribbons correspond to the foot points of the heated
post flare loops, where energetic electrons and downward heat conduction
along the post reconnection loops cause heating and brightening
in the lower atmosphere. In Figure \ref{fig:ribbon_xray}(a) we show
an image of the gradient of temperature increase with height
at the lower boundary, which is a measure of the downward heat conduction
along the loops rooted there.
This is shown at time $t=1.39$ hour, just
after the peak acceleration of the eruption.
It marks the location of the foot points of
strongly heated loops as a result rapid reconnection in the simulation.
We see two curved bright ribbons, which show some similarities
in their paths and locations in relation to the normal magnetic flux
distribution (contours in Figure \ref{fig:ribbon_xray}(a)) as the observed
flare ribbons. For the upper ribbon, its east part extends into the dominant
negative pre-existing sunspot and its west part curves about the emerging
negative polarity region which corresponds to the emerged, fragmented negative
pores to the west of the main negative sunspot in the observation.
The lower ribbon has an overall arc shape and cuts across the main positive
emerging sunspot.
We also plotted field lines with foot points rooted in the bright ribbons
and color the field lines based on temperature as shown in Figure
\ref{fig:ribbon_xray}(b).  We see that these field lines form a sigmoid-shaped
row of heated loops showing an overall morphology similar to that of the
X-ray post flare loop brightening seen in the XRT observation
(Figure \ref{fig:hinode_ribbon_xray}).

\subsection{Evolution of the erupting flux rope \label{sec:erupting_rope}}
By densely tracing field lines inside the erupting cavity (whose cross-section
can be seen in the meridional cross-section plots of density in Figure
\ref{fig:vrmeri_dmeri}), we examine the structure and evolution of the erupting
flux rope.  Figure \ref{fig:erupting_rope_farviews_rev} shows snapshots of
the 3D field lines in the erupting flux rope, with the field lines colored
based on the sign of the north-south component of the magnetic field
$B_z$, with green (purple) indicating positive (negative) $B_z$.
The left column images show the view from the Earth's line-of-sight, and the
right column images shows a side view.
The bottom images show more zoomed out views at a later time
(as indicated by the smaller size of the Sun compared to the other panels).
At the onset of the eruption, the $B_z$ for the top of the flux rope and also
the overlying potential field above the flux rope is positive
or northward (green).  Immediately after the onset of the eruption, the
erupting flux rope rotates counter-clockwise when viewed from above, as
can be seen in Figure \ref{fig:erupting_rope_farviews_rev}.
This direction of rotation is consistent with the left-handed twist of
the pre-eruption flux rope \cite[e.g.][]{Fan:Gibson:2004,Fan:2005}.
By the time the top of the flux rope has reached about $r=2.4 R_{\odot}$
(Figures \ref{fig:erupting_rope_farviews_rev}(c)(f)), the flux rope
has rotated by about $180^{\circ}$ and the $B_z$ of the entire outer
circumference of the flux rope has become negative or southward directed.
Later the erupting flux rope seems to continually expand outward without
further significant change of its orientation, and both the front and flank
of the expanding flux rope maintains southward directed $B_z$ as can be
seen in the zoomed out views of the later evolution shown in
Figures \ref{fig:erupting_rope_farviews_rev}(d)(h).
This orientation of the expanding flux rope, if maintained in the propagation
in the solar wind to the Earth, would explain the southward directed magnetic
field for the front of the magnetic cloud (MC) impacting the Earth
\citep[e.g.][]{Liu:etal:2008}. 
This southward directed $B_z$ for the front of the MC appears opposite to
the northward $B_z$ for the top of the pre-existing flux rope and the overlying
potential field in the CME source region on the Sun.
The large ($\sim 180^{\circ}$) counter-clockwise rotation of the flux rope
achieved during the initial phase of the eruption is consistent with the
large left-handed twist (more than 1 wind of field-line rotation about the
axis) stored in the pre-eruption flux rope.

We find that due to continued magnetic reconnection, the erupting flux
rope evolves to become completely rooted in the ambient pre-existing normal
flux distribution, outside of the emerging bipolar region, as shown in
Figure \ref{fig:footpoints_sig_cusp}(a) where the foot points of
the field lines of the erupting flux rope in
Figures \ref{fig:erupting_rope_farviews_rev}(d)(h) are plotted on
the lower boundary
against the normal magnetic field distribution.
Thus we expect the coronal dimmings, produced by plasma depletion in flux
tubes of the stretched out fields of the CME, to form outside
of the emerging bipolar region, away from the main flare site
\citep[e.g.][]{Gibson:Fan:2008,Attrill:etal:2010,Imada:etal:2011}.
Figure \ref{fig:footpoints_sig_cusp}(b) shows that, a new twisted flux
rope connecting the emerging bipolar region with sigmoid shaped field
lines reforms due to continued flux emergence, under the cusped
loops which have just reconnected in the
overlying vertical current sheet (the purple iso-surface).
The reformation of the flux rope and the associated buildup of
the free magnetic energy after the magnetic energy release of the first
eruption (see upper panel of Figure \ref{fig:emfree_ek_evol}) are
setting up for a second eruption.
This may qualitatively explain the buildup for a second X-class flare
and CME in the same region the following day on December 14.

\section{Discussions and Conclusions}
Improving upon the previous work of F11,
we have carried out an MHD simulation to qualitatively model the
magnetic field evolution of the eruptive flare and CME on 13 December
2006 in the emerging $\delta$-sunspot region AR 10930.
The main improvement compared to F11 is the significantly widened simulation
domain and the inclusion of a much more extended region of the observed
photospheric normal flux distribution in the construction of the
pre-existing coronal potential field.
In this way the simulation can accommodate the wide CME
(see Figure \ref{fig:vrmeri_dmeri}) and better
determine the structure and dynamic evolution of the erupting flux rope
without it being severely constrained by the boundaries immediately after
the onset of the eruption as was the case in F11.

Guided by the observed photospheric magnetic flux emergence
pattern in AR 10930 \citep[e.g.][]{Kubo:etal:2007,
Min:Chae:2009,Ravindra:etal:2011b},
we impose the emergence of an east-west oriented,
left-hand twisted flux rope at the lower boundary.
The resulting flux emergence pattern is such that the positive
emerging polarity corresponds to the observed positive (counter-clockwise)
rotating sunspot emerging against the south end of the pre-existing dominant
negative sunspot, and the negative emerging polarity corresponds to the
collection of the fragmented negative pores observed to emerge to the west
of the $\delta$-spot \citep{Min:Chae:2009}.
As a result of the flux emergence, a twisted coronal flux rope confined by
the pre-existing potential coronal field constructed based on the observed
ambient photospheric magnetic field is built up quasi-statically.  The
resulting pre-eruption coronal magnetic field shows heated, inverse-S shaped
core fields with morphology similar to the bright sigmoid-shaped loops
in the pre-eruption region observed in soft X-ray images by Hinode XRT.

The flux rope is found to erupt after its field line twist about the
axis has reached about 1.2 winds, close to the threshold for the onset
of the helical kink instability, and its upper half of the cross-section
has entered the height where the decline rate of the corresponding
potential field has exceeded the thresh hold for the onset of the torus
instability. The twist of the flux rope at the onset of eruption is
within the measured range of twist transported into the corona based on
the observation of the rotating positive sunspot.
Using a measure of the downward heat conduction flux at the lower boundary
as the proxy for the location of the flare ribbons, we found that
the path and locations of the ribbons in relation to the normal magnetic
flux distribution show qualitative similarities as the observed flare ribbons 
in the lower solar atmosphere observed by the Hinode SOT. Also the field
lines rooted in the bright ribbons form a sigmoid-shaped row of heated loops
that show similar morphology to the observed X-ray post flare loop brightening.

The initial potential field overlying the coronal flux rope and also the
top of the left-hand twisted coronal flux rope of the source region are
both having northward directed (positive) $B_z$ field.
However, immediately after
the onset of the eruption, the erupting flux rope in the
cavity is found to undergo a counter-clockwise rotation (as viewed
from above) by about $180^\circ$ such that its entire front and flanks are
showing southward directed (negative) $B_z$ field (see Figure
\ref{fig:erupting_rope_farviews_rev}). This large rotation of the
erupting flux rope is accomplished in the early phase of the eruption
and later the flux rope shows mainly outward expansion without
further significant change of orientation
(see Figures \ref{fig:erupting_rope_farviews_rev}(c)(d)
and Figures \ref{fig:erupting_rope_farviews_rev}(f)(h)).
The orientation of such an expanding flux rope in the CME ejecta,
if maintained during its propagation in the solar wind, may explain
the southward directed
magnetic field in the front part of the magnetic cloud impacting the
Earth \citep[e.g.][]{Liu:etal:2008}.
The front of the erupting flux rope is found to accelerate to a terminal
speed of about $1500$ km/s, which is still smaller than the observationally
measured range of $1780$ km/s to $3060$ km/s for the CME speed
determined from SOHO Large Angle and Spectrometric Coronagraph (LASCO)
observations \citep[e.g.][]{Ravindra:etal:2010}.
Our simulation also shows that the source region coronal magnetic field
driven by the continued flux emergence is capable of reformation of the
flux rope and a second eruption and hence explaining the observed occurrence
of another eruptive flare on December 14, 2006 from the same region.

\begin{figure}
\centering
\includegraphics[width=0.75\textwidth]{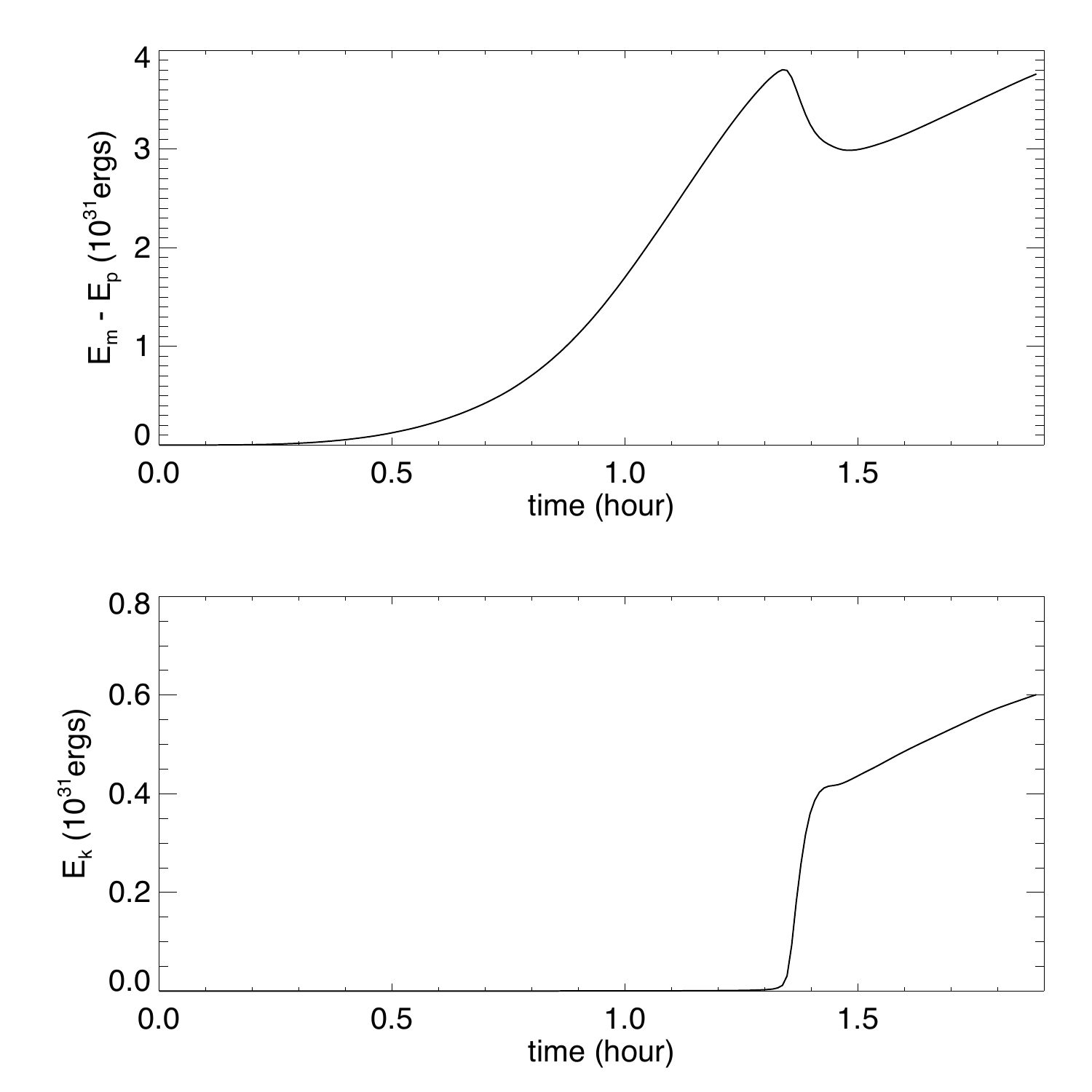}
\caption{Top panel: the evolution of the free magnetic energy,
which is the total magnetic energy $E_m$ minus the energy $E_p$ of
the corresponding potential magnetic field extrapolated using the
current normal magnetic field distribution at the lower boundary.
Bottom panel: the evolution of the kinetic energy.}
\label{fig:emfree_ek_evol}
\end{figure}

\begin{figure}
\centering
\includegraphics[width=0.75\textwidth]{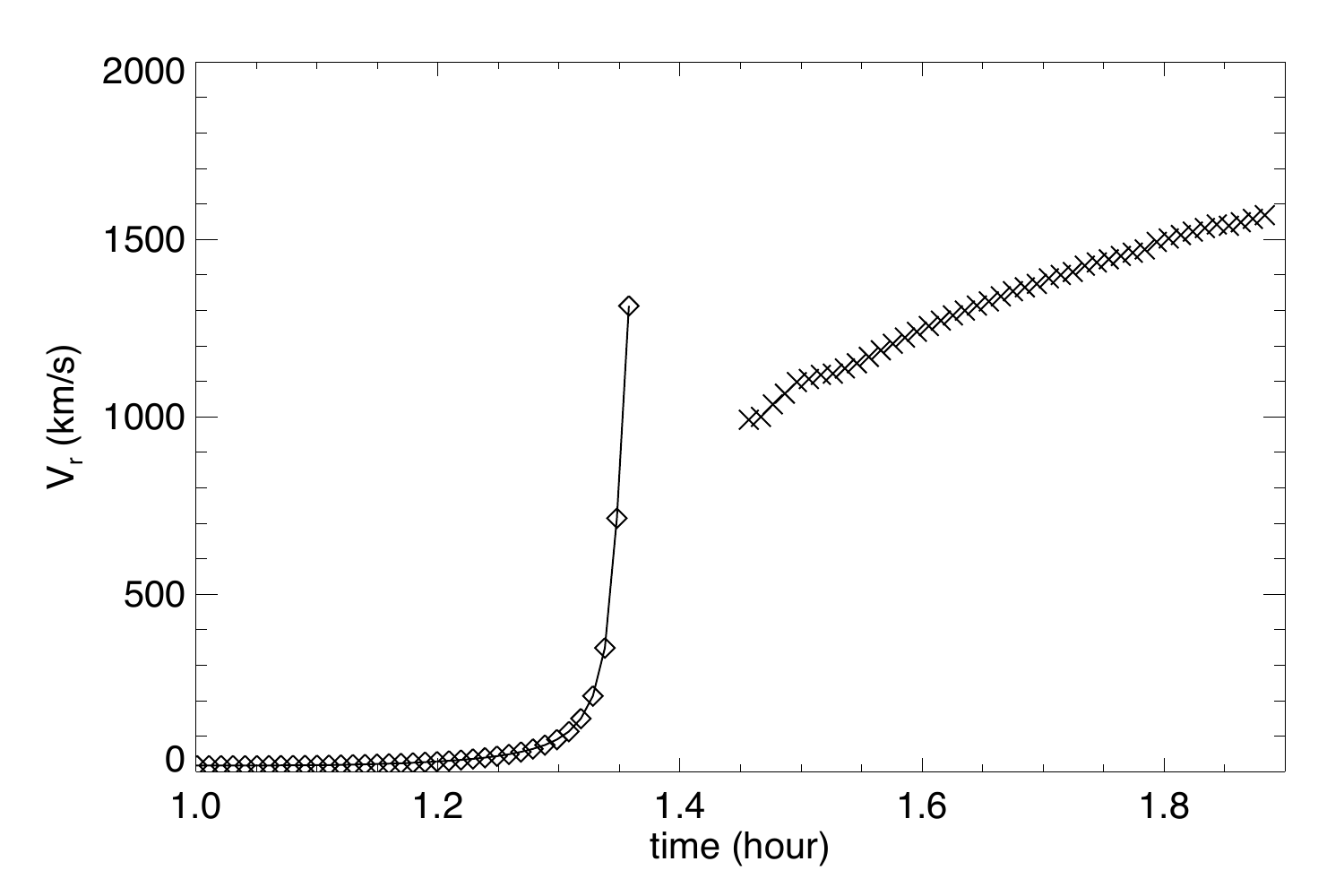}
\caption{The rise velocity of the axis of the coronal flux rope that
forms and then erupts (diamond points) and the rise velocity at the front
of the cavity of the erupting flux rope (cross points).}
\label{fig:vr_axis_sheath}
\end{figure}

\begin{figure}
\centering
\includegraphics[width=0.75\textwidth]{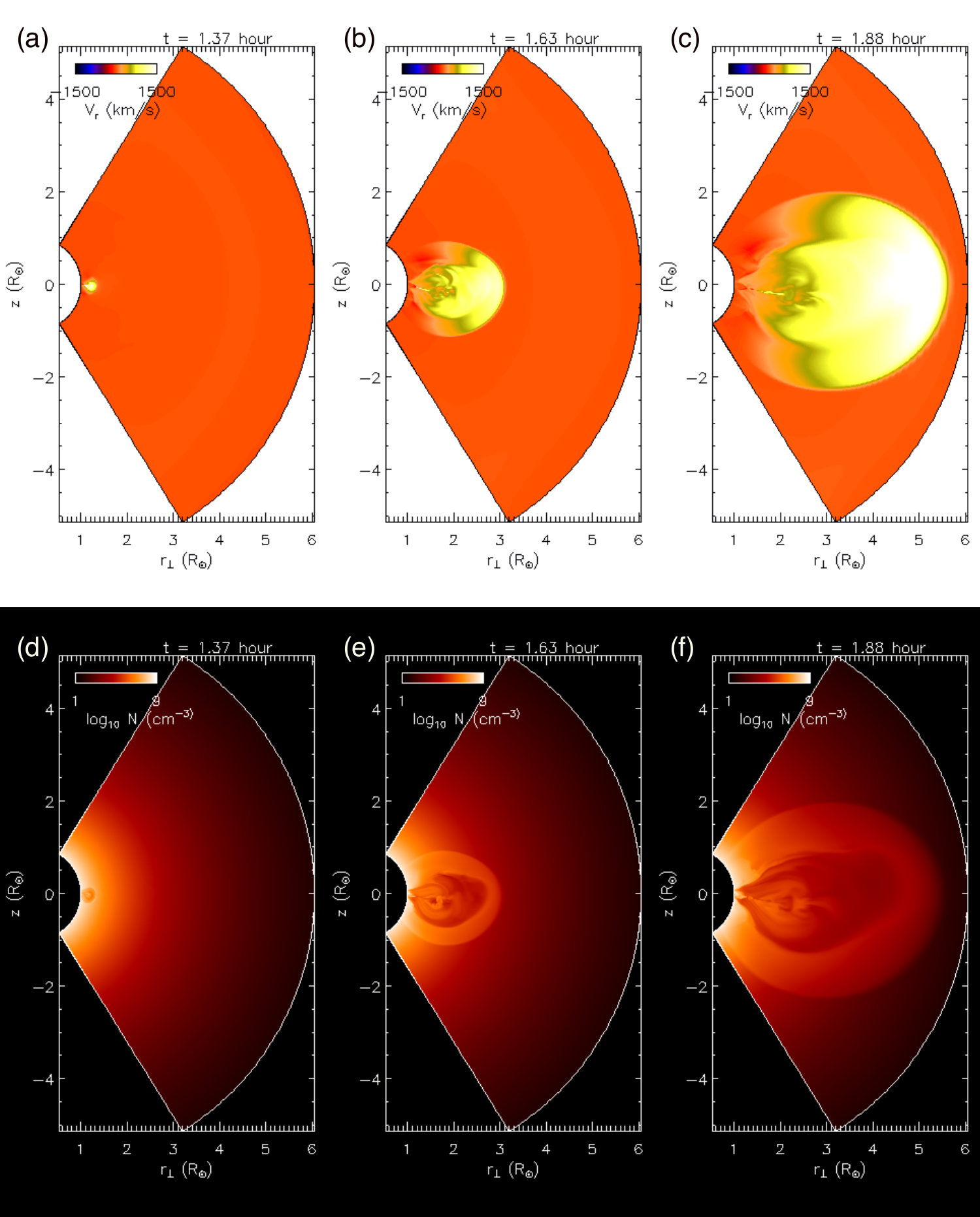}
\caption{(Top panels) Snapshots of the evolution of the radial 
velocity $V_r$ in the central meridional cross-section of the erupting flux 
rope. (Bottom panels) Same as the top panels but showing the density.
A movie of the evolution of both $V_r$ and the density in the central
meridional cross section is available in the online version.}
\label{fig:vrmeri_dmeri}
\end{figure}

\begin{figure}
\centering
\includegraphics[width=0.5\textwidth]{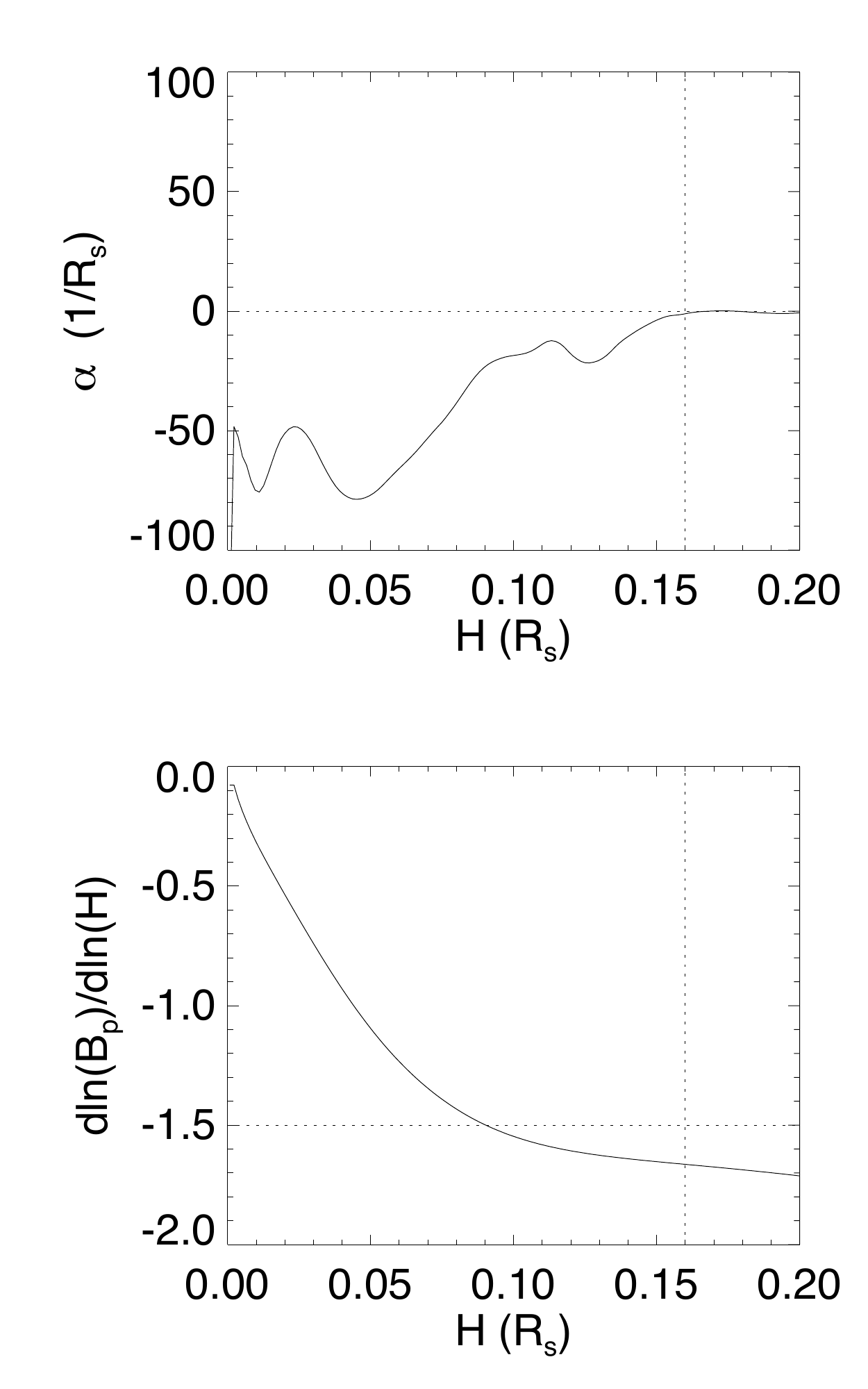}
\caption{Top panel shows the profile of 
$\alpha \equiv (\nabla \times {\bf B}) \cdot {\bf B} / B^2 $ along the
central vertical slice in the central meridional plane across the flux
rope. Bottom panel shows $d \ln (B_p) / d \ln (H)$, the decay rate with
height $H$ of the corresponding potential magnetic field $B_p$
along the same vertical slice. $R_s$ denotes the solar radius.
The value of $\alpha$ is a measure of the
twist rate of the magnetic field, and the height range where $\alpha$ is
significantly negative indicates the height range of the flux rope
cross-section, and thus the vertical dotted lines marks
the top of the flux rope cross section.}
\label{fig:bpdecay_alpha}
\end{figure}

\begin{figure}
\centering
\includegraphics[width=0.75\textwidth]{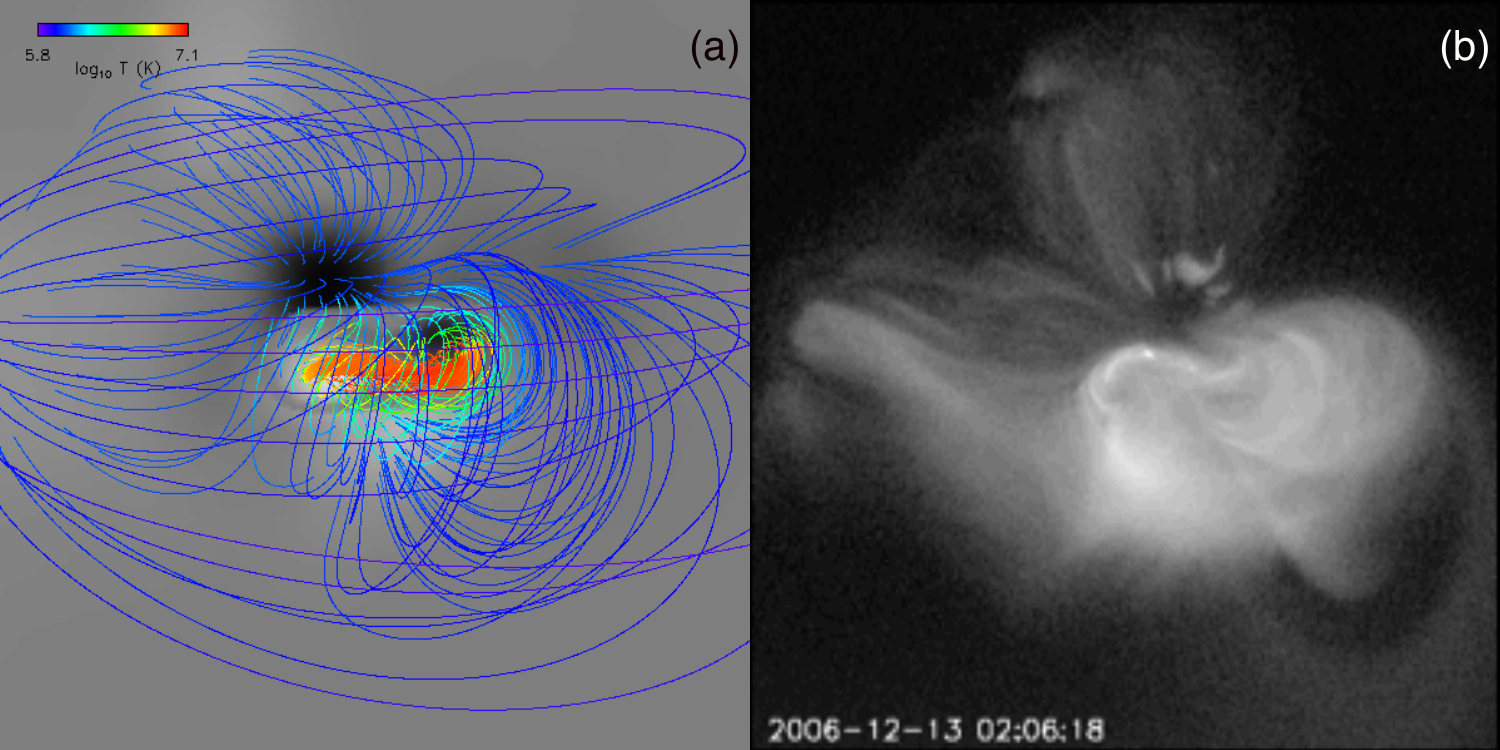}
\caption{3D field lines (colored based on the
temperature) of the coronal magnetic field (with the central heated field
lines densely traced) at a time ($t=1.26$ hour)
just before the onset of the eruption viewed from the Earth's line of sight,
compared with the Hinode XRT image of the
region just prior to the onset of the flare (b).
Panel (b) is reproduced from Figure 8(e) in F11}
\label{fig:cfdl_sigmoidsh}
\end{figure}

\begin{figure}
\centering
\includegraphics[width=0.75\textwidth]{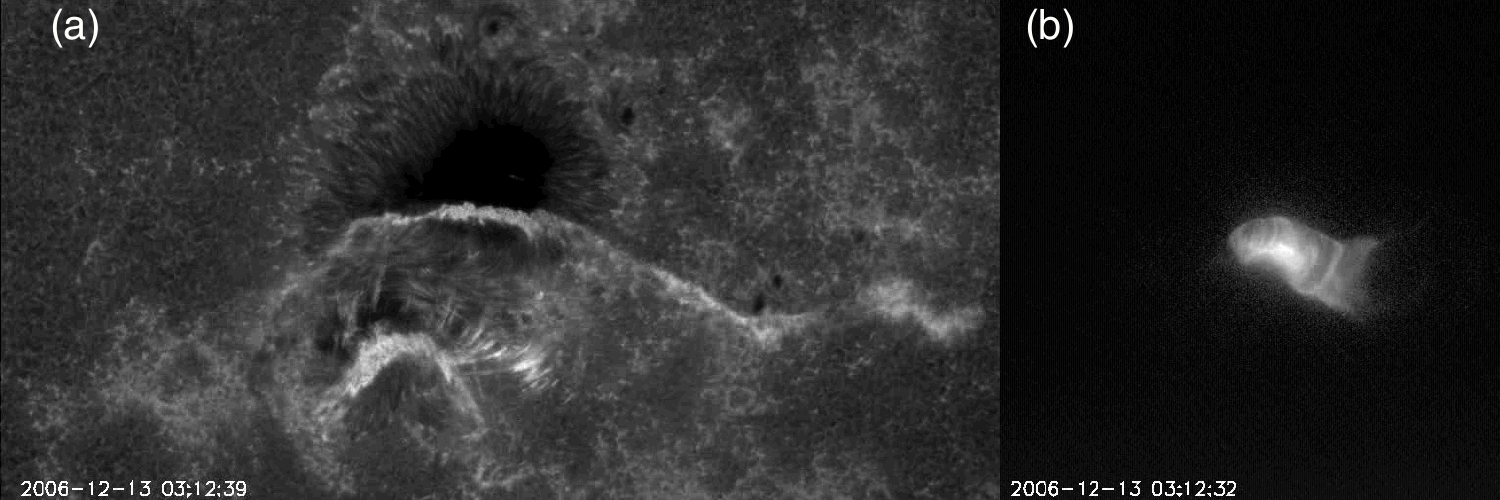}
\caption{(a) Hinode SOT image in Ca II line showing the chromosphere flare
ribbon brightening, and (b) Hinode XRT image showing the soft X-ray post
flare loop brightening.
This figure is reproduced from Figures 9(c)(d) in F11}
\label{fig:hinode_ribbon_xray}
\end{figure}

\begin{figure}
\centering
\includegraphics[width=0.75\textwidth]{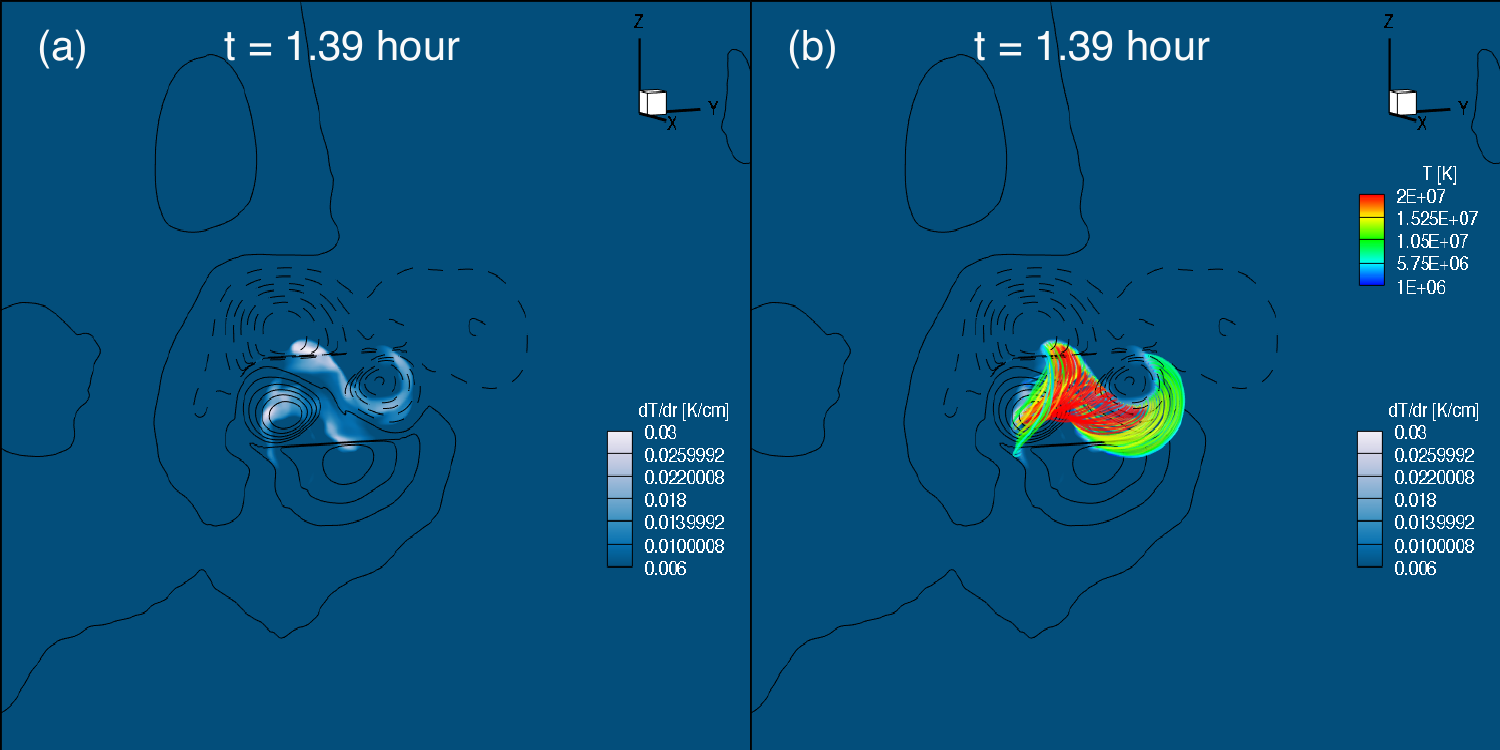}
\caption{(a) Image showing the vertical gradient of temperature increase 
on the lower boundary overlaid with the line contours showing the normal
magnetic field with solid contours (dashed contours) representing
positive (negative) magnetic polarity. (b) Same as (a) but also showing
field lines with foot points rooted in the bright ribbons in (a).
The field lines are colored based on temperature.}
\label{fig:ribbon_xray}
\end{figure}

\begin{figure}
\centering
\includegraphics[width=0.55\textwidth]{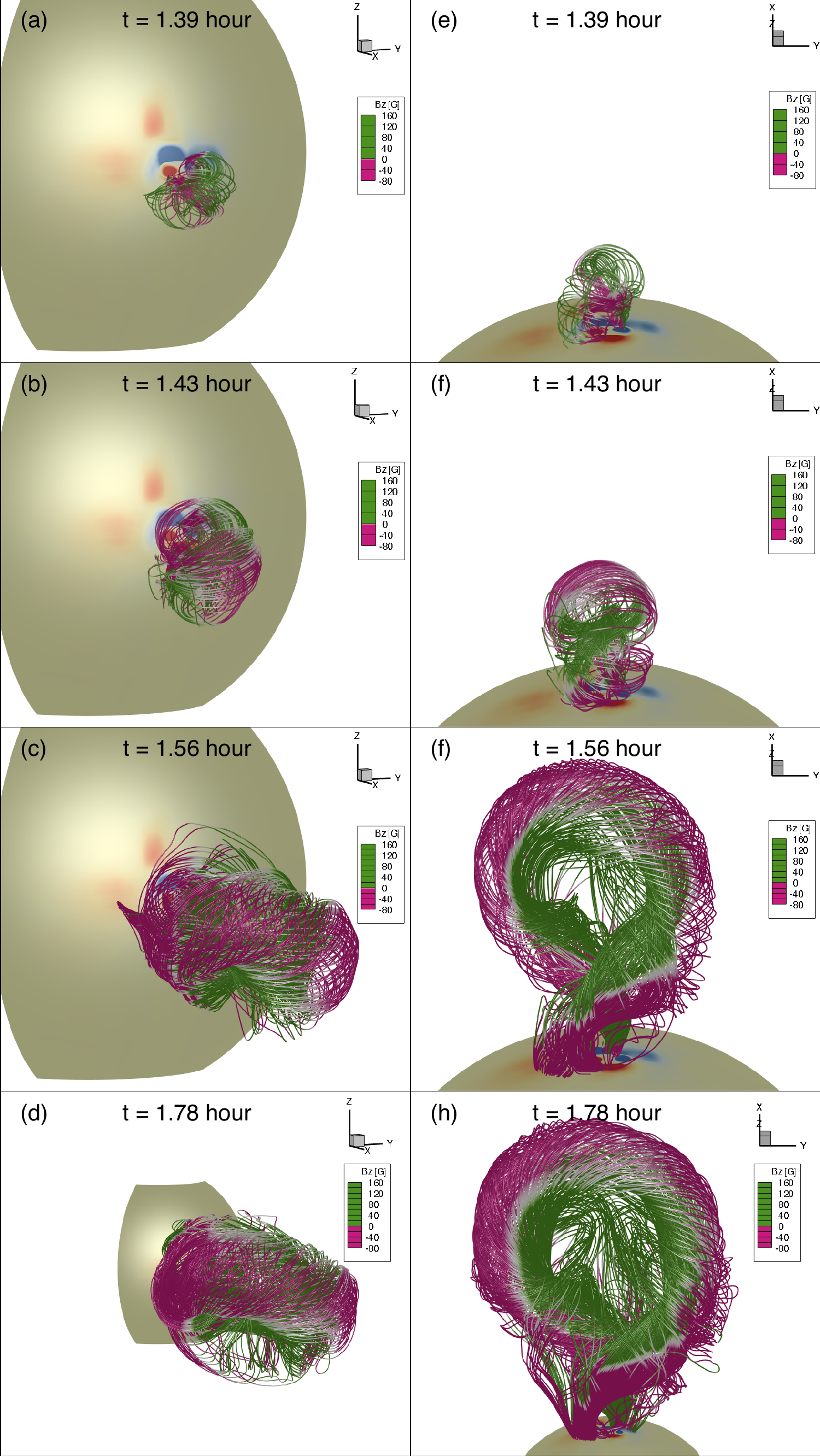}
\caption{Snapshots of the 3D field lines of the erupting flux rope, with
the left column images showing a view from the Earth's line-of-sight and the
right column images showing a side view.  The field lines are colored based
on the sign of the north-south component of the magnetic field $B_z$. The
color on the sphere shows the normal magnetic field strength on the lower
boundary with red (blue) corresponding to positive (negative) field.
The bottom images are more zoomed out views as indicated by the smaller size of
the Sun}
\label{fig:erupting_rope_farviews_rev}
\end{figure}

\begin{figure}
\centering
\includegraphics[width=1.\textwidth]{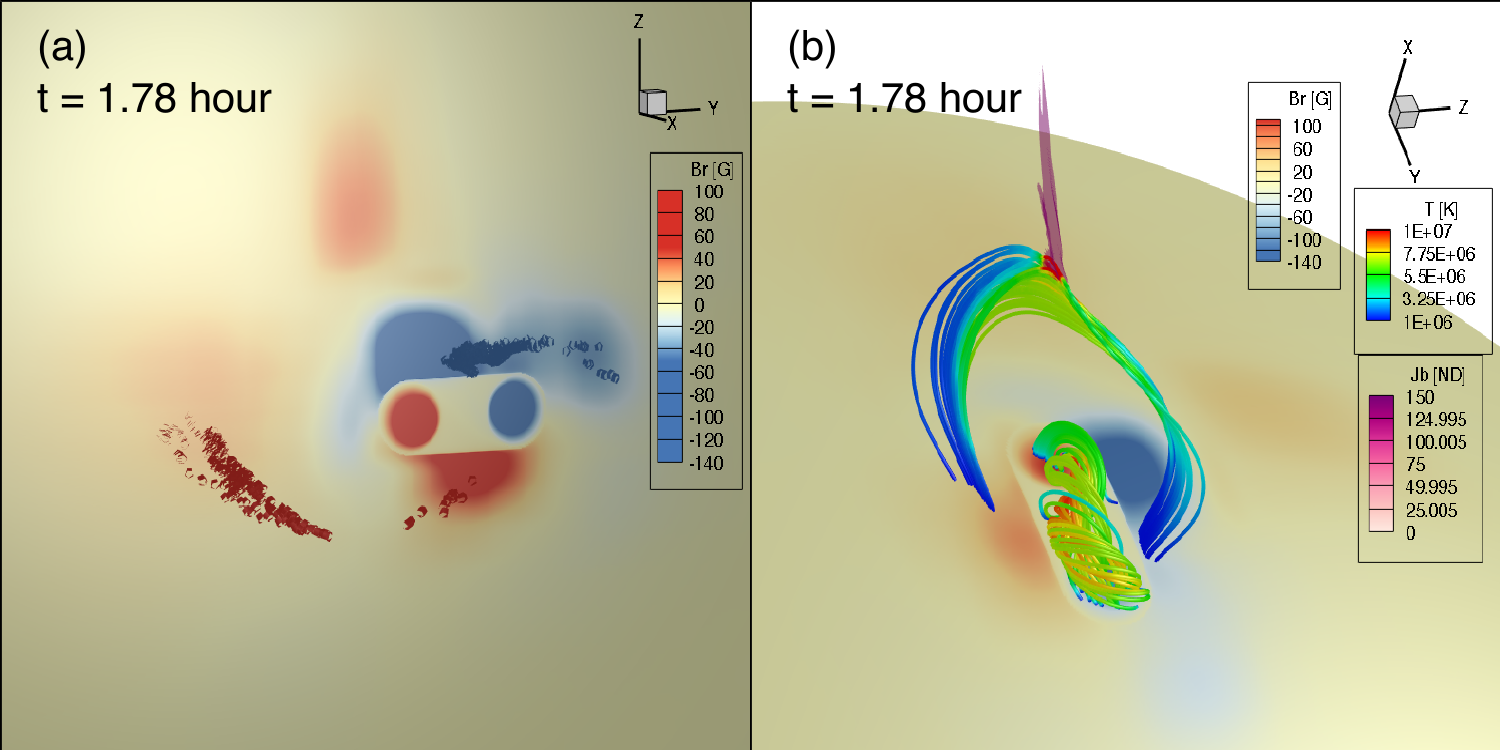}
\caption{(a) The foot points of the field lines of the erupting flux rope in
Figures \ref{fig:erupting_rope_farviews_rev}(d)(h) plotted on the lower
boundary against the normal magnetic field $B_r$.
The foot points are colored based on the polarity region in which they are
located, red for positive normal field and blue for negative.
(b) Selected field lines (colored based on the temperature) showing the
field lines of the reformed twisted flux rope connecting the emerging bipolar
region under a set of cusped field lines just reconnected in the overlying
vertical current sheet
(represented by the purple iso-surface of
$J_b \equiv |\nabla \times {\bf B} | / B = 1/(5 dr)$, where $dr$ is
the radial grid size)}.
\label{fig:footpoints_sig_cusp}
\end{figure}

Our MHD simulation driven by the emergence of an east-west oriented magnetic
flux rope is aimed to model qualitatively the structure and topology of the
magnetic field for both the source region and the CME ejecta of
the eruptive flare on December 13, 2006.
There are many severe limitations of the model. The smoothing of the 
observed photospheric field to reduce the peak field strength at the lower
boundary to under $200$ G, due to numerical constraints, greatly under estimates
the free magnetic energy buildup (about $4 \times 10^{31}$ ergs) and
release ($\sim 10^{31}$ ergs) for the resulting CME
(Figure \ref{fig:emfree_ek_evol}), compared to the observational
estimates (on the order of $10^{32}$ ergs) for the
CME energy \citep[e.g.][]{Schrijver:etal:2008,Ravindra:etal:2010}.
This would lead to an under estimate of the CME speed.
Also, our simulation assumes an initial
potential field without an ambient solar wind.  All of these
simplifications can affect the resulting CME speed and energetics.
Further improvement of the model with more realistic lower
boundary conditions and the inclusion of an ambient solar wind
with more realistic treatment of the thermodynamics are needed to
achieve a quantitative description of the event.
Progress is being made in using high-cadence vector magnetic
field and Doppler velocity observations by the Helioseismic and
Magnetic Imager (HMI) of the Solar Dynamics Observatory (SDO) to
infer the electric field evolution at the photosphre of
flare/CME productive active regions \citep[e.g.][]{Kazachenko:etal:2014,
Kazachenko:etal:2015}. Such observationally inferred electric fields may be
used for constructing more realistic lower boundary driving conditions of
flux emergence for the MHD simulations of the CME events.

\acknowledgments
This work has been supported in part by the NASA HSR grant NNX13AK54G,
and the Air Force Office of Scientific Research grant FA9550-15-1-0030 to NCAR.
NCAR is sponsored by the National Science Foundation. The numerical
simulations were carried out on the Yellowstone supercomputer of NWSC/NCAR
under the NCAR Strategic Capability computing project NHAO0001, and also
on the Pleiades supercomputer at the NASA Advanced Supercomputing Division
under project GID s01368 and the Discover supercomputer at NASA Center for
Climate Simulation under the project GID s01362.

\clearpage

\clearpage

\clearpage

\clearpage

\clearpage

\clearpage

\clearpage

\clearpage

\clearpage

\clearpage

\clearpage

\clearpage

\end{document}